# The phase-functions method and full cross-section of nucleon-nucleon scattering

**V. I. Zhaba**

Uzhgorod National University, Department of Theoretical Physics,
54, Voloshyna St., Uzhgorod, UA-88000, Ukraine

**Abstract**

For calculation of the single-channel nucleon-nucleon scattering a phase-functions method has been considered. Using a phase-functions method the following phase shifts of a nucleon-nucleon scattering are calculated numerically: *nn* ($^1S_0$-, $^3P_0$-, $^3P_1$-, $^1D_2$-, $^3F_3$- state), *pp* ($^1S_0$-, $^3P_0$-, $^3P_1$-, $^1D_2$-state) and *np* ($^1S_0$-, $^1P_1$-, $^3P_0$-, $^3P_1$-, $^1D_2$-, $^3D_2$- state). The calculations has been performed using realistic nucleon-nucleon potentials Nijmegen groups (NijmI, NijmII, Reid93) and potential Argonne v18. Obtained phase shifts are in good agreement with the results obtained in the framework of other methods. Using the obtained phase shifts we have calculated the full cross-section. Our results are in good agreement with those obtained by using known phases published in literature. The odds between calculations depending on a computational method of phases of scattering makes: 0,2-6,3% for *pp*- and 0,1-5,3% for *np*- scatterings (NijmI, NijmII), 0,1-4,1% for *pp*- and 0,1-0,4% for *np*- scatterings (Reid93), no more than 4,5% (Argonne v18).

## 1. Introduction

From the experimentally observed values of the scattering cross sections and energies of transitions in the first receive queue information about the phase and amplitude of the scattering wave function, which is the main object of study in the standard approach. In other words, in the experiment observed not the wave functions, and their changes caused by the interaction [1]. It is therefore of interest to obtain the equation directly connect the phase and amplitude of the scattering potential, finding the wave functions.

Exact solution of the scattering problem to calculate the scattering phase is possible only for individual phenomenological potentials. When realistic potentials are used, the phase of the scattering are calculated approximately. This is due to the use of physical approximations or numerical calculation. The influence of the choice of numerical algorithm for the solution of scattering problems listed in Ref. [2, 3].

The methods of solving the Schrödinger equation for the purpose of obtaining the scattering phases are: the method of successive approximations, Born approximation, phase-functions method and others.

This paper is devoted to calculation of the phase shifts and total cross sections of nucleon-nucleon scattering for the modern realistic phenomenological nucleon-nucleonic the Nijmegen groups (NijmI, NijmII, Reid93) and Argonne v18 potentials by using the phase function method (PFM).

## 2. The phase-functions method

Consider a typical example of the reaction: scattering a spin-free particle with certain specific values of energy $E$ (or a wave number $k$: $k^2 = \frac{2mE}{\hbar^2}$) and the orbital moment l in a spherical symmetric potential $V(r)$. The Schrödinger equation for the corresponding radial wave function $u_l(r)$ has the form [1]:

$$u''_l(r) + \left(k^2 - \frac{l(l+1)}{r^2} - U(r)\right) u_l(r) = 0, \qquad (1)$$

where $U(r) = \frac{2m}{\hbar^2} V(r)$ - the renormalized interaction potential, $m$ - the reduced mass.

PFM is a special method of solution of the radial schrödinger equation (1), which is a linear differential equation of second order. It is quite easy to obtain the scattering phase, because this method do not need to calculate in a wide range of radial wave functions of scattering problems and then they find the asymptotics of these phases [1, 4].

Two linearly independent solutions of the free Schrödinger equation (1) (with $U \equiv 0$) are known Riccati-Bessel functions $j_l(kr)$ and $n_l(kr)$. The free motion corresponds to the only regular at the point $r$=0, the solution $j_l(kr)$, so in this case, asymptotically for large values of r the solution takes the form

$$u_l(r) \approx const \cdot \sin(kr - l\pi/2). \qquad (2)$$

The potential leads to the fact that it is now in the area where the potential $U(r)$ vanishes, the wave function includes an additive irregular solutions of the free equation $n_l(kr)$. The degree of this supplement, quantitatively describes the effect of the interaction is the scattering phase shift $\delta_l$:

$$u_l(r) \approx const \cdot \left[ j_l(kr) - tg\delta_l \cdot n_l(kr) \right],$$

$$u_l(r) \to const \cdot \sin(kr - l\pi/2 + \delta_l), \quad r \to \infty.$$

A set of phase shifts $\delta_l$ for different partial waves determines the angular distribution and total cross section for scattering. Therefore, an important problem in the theory of potential scattering is the search for values of $\delta_l$ for the given values of the potential $U(r)$, the orbital moment $l$ and energy $E$. The standard method of calculating the scattering phase is a solution of the Schrödinger equation (1) with the asymptotic boundary condition. PFM is the transition from equation of Schrödinger to the equation for the phase function. For this change [1, 4]:

$$u_l(r) = A_l(r)\left[\cos\delta_l(r) \cdot j_l(kr) - \sin\delta_l(r) \cdot n_l(kr)\right]. \qquad (3)$$

Introduced two new functions $\delta_l(r)$ and $A_l(r)$ are the corresponding scattering phase and normalization constants (amplitudes) of wave functions for scattering on a sequence of truncated potentials. $\delta_l(r)$ and $A_l(r)$ are called according to their physical content phase and amplitude function. The term “phase function” was first used in paper Morse and Allis [5]. Equations for phase and amplitude functions with the initial conditions are:

$$\delta'_l(r) = -\frac{1}{k} U(r)\left[\cos\delta_l(r) \cdot j_l(kr) - \sin\delta_l(r) \cdot n_l(kr)\right]^2, \ \delta_l(0) = 0; \qquad (4)$$

$$A'_l(r) = -\frac{1}{k} A_l(r) U(r)\left[\cos\delta_l(r) \cdot j_l(kr) - \sin\delta_l(r) \cdot n_l(kr)\right] \times$$
$$\times \left[\sin\delta_l(r) \cdot j_l(kr) + \cos\delta_l(r) \cdot n_l(kr)\right], \ A_l(0) = 1. \qquad (5)$$

The phase equation (4) for the first time has been obtained Drukarev [6], and then independent by Bergmann [7], Kynch [8, 9], Olsson [10] and Calogero [11]. Except for the shape of an entry (4), other common shapes of the phase equation are possible also, that considered Calogero and Ravenhall [12], Franchetti [13], Dashen [14], Swan [15], Zemach [16].

Note the advantages of the PFM approach to calculate phases compared to the standard method based on the consideration of the Schrödinger equation for the wave function [1]. The fact that the phase equation is of first order (though nonlinear), for easy programming and calculation on the computer. In addition, this reduces the number of operations and computation time. Not oscillating, but more monotonous behavior of the phase function allows for calculations with greater precision and facilitates the estimation of the uncertainties of the results. Another important advantage of this method is the possibility of building within its framework, new algorithms calculate not only phase shifts but also the partial and full scattering amplitudes, *S*-matrix, scattering lengths, effective radii and other parameters of the scattering [1].

The special case of the phase equation (4) at $l=0$ has been used at examination of problem *S*-scattering of slow electrons on atoms [5]. PFM appeared convenient at a solution of many problems atomic and a nuclear physics.

And if the last century PFM was used more often. But all at contributors remains particular interest to its application in calculations. For example, the quasilinearization method [17] with a PFM gives excellent results when applied to computation of ground and excited bound state energies and wave functions for a variety of the potentials. In Ref. [18] a generalization of the phase function method to the problems of acoustic wave scattering on continuous medium inhomogeneities is proposed.

## 3. Potentials of nucleon-nucleon interaction

Why it was chosen to calculate phase shifts using potentials Nijmegen group (NijmI, NijmII, Nijm93, Reid93 [19]) and Argonne v18 [20] potentials? The parameters of potential models optimized in such a way that minimized the value of $\chi^2$ in the direct fit of the data. The first improvement potential Nijm78 [21] was started in the early 1990th years. An improved version of the potential had the name Nijm92pp, since it was updated for 1787 *pp* data. For potential Nijm92pp the value of $\chi^2/N_{pp}$ was 1.4. The following improvement Nijm78 potential for np data gave Nijm93 model: $\chi^2/N_{pp}$ =1.8 for pp 1787 and $\chi^2/N_{nn}$=1,9 for 2514 *np* data, i.e. $\chi^2/N_{data}$=1.87 a. For the Nijm I and NijmII potential the value of $\chi^2/N_{data}$=1,03. The original capacity of the Reid68 potential [22] was parameterized on the basis of the phase analysis Nijmegen group and got the name Reid93. Parameterization was conducted for 50 parameters $A_{ij}$ and $B_{ij}$ of the potential, with $\chi^2/N_{data}$=1,03 [19]. The structure and writing the Reid93 potential is quite cumbersome.

The Argonne v18 potential [20] with 40 regularized parameters gives magnitude $\chi^2/N_{data}$=1,09 for 4301 *pp* and *np* the data in the field of energies 0-350 MeV. For a potential of CD-Bonn [23] magnitude $\chi^2/N_{data}$ makes 1,01 for 2932 *pp* the data and 1,02 for 3058 *np* the data. Such potentials as Hamada-Johnston-62, Yale group potentials, Reid68, UrbanaV14, etc. have major values $\chi^2$, as parameterized on the foundation of narrower energy interval.

So, the Nijmegen groups potentials and Argonne v18 potential is one of those realistic phenomenological potentials who are more best feature nucleon-nucleon interaction. By the way, at calculations of phases of scattering it is necessary to take into account singularities of a potential between nucleon interactions.

## 4. Calculations of phase shifts and full cross-section

Spin states for a neutron-proton system are represented as $^{2S+1}L_J$, where $L$ - the moment of a system (a value of orbital moment $L=0$; 1; 2; 3; 4;… answer S-, P-, D-, F-, G,… - states); $S$ - a spin of a system; $J$ - the complete moment of a system; $P=(-1)^L$ - parity. For *pp*-and *nn*-systems spin states will be $^1S_0$-, $^3P_0$-, $^3P_1$-, $^1D_2$-, $^3F_3$- states. For *np*-systems spin states will be $^1S_0$-, $^1P_1$-, $^3P_0$-, $^3P_1$-, $^1D_2$-, $^3D_2$- states.

The phase-functions method numerically obtains phase shifts of nucleon-nucleon scattering for potentials Nijmegen groups (NijmI, NijmII, Reid93) and potential Argonne v18: *nn* ($^1S_0$-, $^3P_0$-, $^3P_1$-, $^1D_2$-, $^3F_3$- states), *pp* ($^1S_0$-, $^3P_0$-, $^3P_1$-, $^1D_2$- states) and *np* ($^1S_0$-, $^1P_1$-, $^3P_0$-, $^3P_1$-, $^1D_2$-, $^3D_2$- states). Masses of nucleons are chosen such: $M_p$=938,27231 MeV; $M_n$=939,56563 MeV. The numerical method of a solution of the phase equation (4) - a Runge-Kutta method of the fourth order [24] has been elected. Phase shifts were at an exit of phase function $\delta_l(r)$ on an asymptotics at $r$=10 fm. The gamut of energies made 1-350 MeV. Calculations compared with outcomes of other operations: for the partial-wave analysis (PWA) [19], for potentials NijmI, NijmII and Nijm93 [19], Argonne v18 (Av18) [20] and CD-Bonn [23].

As in the scientific literature there were missing tabulared values of phase shifts *nn*-scatterings for potentials of the Nijmegen group, designed on PFM phase shifts compared with phases for potentials Argonne v18 and CD-Bonn.

If to compare the phase shifts of nucleon-nucleon scattering designed for same potentials Nijmegen groups (for *pp*- and *np*- scatterings) different methods - on the basis of a solution of a Schrödinger equations (see Ref. [19]) and on foundation PFM (outcomes of the given paper) it is possible to draw a conclusion, that the discrepancy between outcomes makes no more than two percents.

Comparison of outcomes of calculations of phase shifts for potentials Nijmegen groups, obtained with help PFM, and phase shifts for other potential models (Argonne v18 [20] and CD-Bonn [23]) and for the partial-wave analysis specifies that the deviation between these data makes up to five percents.

The odds between calculations of phase shifts for potential Argonne v18 depending on a method of their deriving makes some percents. Some outcomes of numerical calculations of phase shifts are indicated in [25].

Alongside with phase shifts in problems of scattering it is necessary to deal with scattering amplitudes, devices *S*- matrix and a lot of other parameters. On known phases of scattering calculate the complete amplitude, the full cross-section and a partial scattering amplitude accordingly [1]

$$F(\theta) = \frac{1}{k}\sum_{l=0}^{\infty}(2l+1)e^{i\delta_l}\sin\delta_l P_l(\cos\theta), \tag{6}$$

$$\sigma = \frac{4\pi}{k^2}\sum_{l=0}^{\infty}(2l+1)\sin^2\delta_l , \tag{7}$$

$$f_l = \frac{1}{k}e^{i\delta_l}\sin\delta_l , \tag{8}$$

where $P_l(\cos\theta)$ - polynomials of Legendre, $\theta$ - polarization angle.

Depending on a computational method the modification of a phase of scattering reduces in respective alteration of above-stated magnitudes $F(\theta)$, $\sigma$, $f_l$. For example, for $^1S_0$- state *np*- system a modification of a phase of scattering $\delta_l$ on 1-2% gives minor alteration (no more than 5%) to the real and imaginary part of partial amplitude $f_l$.

Comparison between magnitudes of the full cross-section of nucleon-nucleon scattering $\sigma$ (in $fm^{-2}$), designed on phase shifts from Ref. [19, 20] is carried out and to phase shifts agrees PFM (outcomes of the given paper) for potential NijmI, NijmII, Reid93 and Argonne v18. Outcomes of calculations of the full cross-section of scattering (7) are reduced in Tables 1-3. The odds between calculations depending on a computational method of phases of scattering makes: 0,2-6,3% for *pp*- and 0,1-5,3% for *np*- scatterings (NijmI, NijmII), 0,1-4,1% for *pp*- and 0,1-0,4% for *np*- scatterings (Reid93), no more than 4,5% (Argonne v18). Association of the full cross-section of scattering $\sigma$ from phase shifts $\delta_l$ is determined component $\sin^2\delta_l$ of the total (7).

## Conclusions

The known phase-functions method is considered for a problem of single-channel nucleon-nucleon scattering.

For the first time by a method of phase functions are calculate *nn*, *pp* and *np* phase shifts for the relevant spin configurations for nucleon-nucleon potentials Nijmegen groups (NijmI, NijmII, Reid93) and potential Argonne v18. Numerical the obtained phase shifts well agree with outcomes of original Ref. [19, 20] for the same potentials (the deviation makes no more than two percents). Also compared outcomes of calculations of phase shifts with help PFM with phase shifts for other potential models and for the partial-wave analysis: the deviation between these data makes up to five percents.

Despite of labour-intensive process of numerical calculations, on the obtained phase shifts on PFM the full cross-section of nucleon-nucleon scattering $\sigma$ is designed, and also its comparison with magnitude of the complete cut obtained on phase shifts from papers [19, 20] is made. It is obtained also a partial scattering amplitude $f_l$ for $^1S_0$- state *np*- system. A value $\sigma$ and $f_l$ insignificantly differ from the magnitudes obtained on known phases from paper for the indicated potential Reid93 between nucleon of interaction.

**Table 1.** The full cross-section of scattering (NijmI and NijmII)

| $T_{lab}$, MeV | $\sigma_{pp}$ [19] | $\sigma_{pp}$ PFM | $\sigma_{np}$ [19] | $\sigma_{np}$ PFM | $\sigma_{pp}$ [19] | $\sigma_{pp}$ PFM | $\sigma_{np}$ [19] | $\sigma_{np}$ PFM |
|---|---|---|---|---|---|---|---|---|
| | NijmI | NijmI | NijmI | NijmI | NijmII | NijmII | NijmII | NijmII |
| 1 | 76,43 | 77,05 | 203,57 | 204,73 | 76,47 | 77,34 | 203,50 | 204,55 |
| 5 | 175,12 | 173,38 | 210,95 | 211,33 | 175,24 | 174,49 | 210,66 | 210,93 |
| 10 | 180,23 | 180,60 | 202,53 | 202,88 | 180,30 | 181,38 | 202,08 | 202,24 |
| 25 | 170,38 | 170,97 | 194,36 | 194,56 | 170,19 | 173,67 | 193,60 | 194,58 |
| 50 | 151,62 | 150,59 | 209,33 | 210,49 | 151,24 | 152,07 | 208,84 | 211,44 |
| 100 | 114,58 | 115,91 | 279,38 | 281,15 | 114,87 | 117,78 | 279,62 | 278,10 |
| 150 | 105,24 | 105,44 | 369,83 | 375,39 | 105,63 | 107,58 | 368,79 | 374,61 |
| 200 | 127,40 | 122,31 | 462,99 | 458,01 | 126,69 | 128,36 | 459,88 | 462,29 |
| 250 | 173,87 | 171,39 | 553,32 | 538,46 | 171,53 | 169,17 | 549,59 | 545,88 |
| 300 | 237,09 | 232,03 | 638,68 | 617,33 | 232,80 | 224,10 | 637,06 | 627,45 |
| 350 | 310,94 | 299,13 | 718,04 | 707,62 | 304,82 | 305,31 | 721,31 | 718,38 |

**Table 2.** The full cross-section of scattering (Reid93)

| $T_{lab}$, MeV | $\sigma_{nn}$ PFM | $\sigma_{pp}$ [19] | $\sigma_{pp}$ PFM | $\sigma_{np}$ [19] | $\sigma_{np}$ PFM |
|---|---|---|---|---|---|
| 1 | 181,96 | 78,43 | 79,59 | 202,74 | 202,78 |
| 5 | 196,96 | 174,98 | 179,56 | 209,07 | 209,09 |
| 10 | 189,00 | 180,07 | 179,39 | 199,94 | 200,03 |
| 25 | 172,29 | 170,48 | 170,68 | 191,70 | 191,91 |
| 50 | 151,86 | 152,72 | 150,98 | 209,81 | 210,54 |
| 100 | 114,90 | 116,28 | 115,35 | 283,50 | 284,34 |
| 150 | 106,17 | 105,73 | 107,21 | 370,21 | 371,07 |
| 200 | 129,45 | 126,06 | 129,11 | 457,23 | 458,20 |
| 250 | 176,81 | 171,13 | 175,20 | 543,29 | 544,54 |
| 300 | 241,17 | 233,35 | 238,73 | 628,32 | 629,44 |
| 350 | 316,97 | 307,12 | 313,58 | 711,73 | 713,03 |

**Table 3.** The full cross-section of scattering (Argonne v18)

| $T_{lab}$, MeV | $\sigma_{nn}$ [20] | $\sigma_{nn}$ PFM | $\sigma_{pp}$ [20] | $\sigma_{pp}$ PFM | $\sigma_{np}$ [20] | $\sigma_{np}$ PFM |
|---|---|---|---|---|---|---|
| 1 | 183,58 | 183,58 | 80,14 | 79,80 | 203,23 | 203,27 |
| 5 | 199,03 | 199,01 | 177,79 | 177,73 | 209,47 | 210,17 |
| 10 | 190,61 | 190,56 | 181,54 | 181,05 | 201,42 | 201,53 |
| 25 | 173,69 | 173,50 | 172,87 | 172,11 | 193,12 | 193,83 |
| 50 | 152,73 | 152,40 | 151,46 | 151,70 | 207,67 | 208,81 |
| 100 | 115,73 | 115,47 | 113,23 | 113,55 | 273,40 | 274,46 |
| 150 | 109,16 | 108,80 | 105,11 | 106,16 | 357,63 | 358,54 |
| 200 | 134,78 | 134,53 | 129,53 | 130,06 | 446,39 | 447,31 |
| 250 | 185,18 | 184,81 | 178,89 | 178,23 | 537,10 | 538,70 |
| 300 | 252,57 | 252,38 | 245,13 | 243,97 | 629,36 | 631,28 |
| 350 | 331,26 | 331,11 | 322,08 | 320,85 | 722,59 | 723,55 |